\documentclass{article}

\usepackage{PRIMEarxiv}

\usepackage[utf8]{inputenc} 
\usepackage[T1]{fontenc}    
\usepackage{hyperref}       
\usepackage{url}            
\usepackage{booktabs}       
\usepackage{amsfonts}       
\usepackage{nicefrac}       
\usepackage{microtype}      
\usepackage{yhmath}
\usepackage{lipsum}
\usepackage{fancyhdr}       
\usepackage{graphicx}       
\usepackage{multirow}       
\graphicspath{{media/}}     
\usepackage{siunitx}

\title{Microfluidics for Hydrodynamics Investigations of Sand Dollar Larvae
\thanks{\textit{\underline{Citation}}: 
\textbf{Chen, et al., Microfluidics for Hydrodynamics Investigations of Sand Dollar Larvae ,  DOI:000000/11111.}}
}

\author
{Wesley A. Chen,$^{1,\#}$ Bryant A. Lopez,$^{1,\#}$  Haley B. Obenshain,$^{1,\#}$ 
Moses Villeda,$^{1,\#}$\\ Brian T. Le,$^{1}$ Brenda AAB. Ametepe,$^{1}$ Ariana Lee,$^{2}$ Douglas A. Pace,$^{2,\star}$ Siavash Ahrar$^{1,\star}$\\
\\
\normalsize{$^{1}$Department of Biomedical Engineering, CSU Long Beach, CA, USA.}\\
\normalsize{$^{2}$ Department of Biological Sciences, CSU Long Beach, CA, USA.}\\
\normalsize{\# authors contributed equally and ${\star}$ to whom correspondence should be addressed.}\\
\normalsize{E-mail contacts: Douglas.Pace@csulb.edu and Siavash.Ahrar@csulb.edu}
}

\begin{document}
\maketitle

\newenvironment{sciabstract}{%
\begin{quote} }
{\end{quote}}

\begin{sciabstract}
The life cycle of most marine invertebrates includes a planktonic larval stage before metamorphosis to bottom-dwelling adulthood. During the larval stage, ciliary-mediated activity enables feeding (capture unicellular algae) and transport of materials (oxygen) required for the larva's growth, development, and successful metamorphosis. Investigating the underlying hydrodynamics of these behaviors is valuable for addressing fundamental biological questions (e.g., phenotypic plasticity) and advancing engineering applications (biomimetic design). In this work, we combined microfluidics and fluorescence microscopy as a miniaturized PIV (mPIV) approach to study ciliary-mediated hydrodynamics during suspension feeding in sand dollar larvae (\emph{Dendraster excentricus}). First, we confirmed the approach's feasibility by examining the underlying hydrodynamics (ciliary-mediated vortex patterns) for low- and high-fed larvae. Next, ciliary hydrodynamics were tracked from 11 days post-fertilization (DPF) to 20 DPF for 21 low-fed larvae. Microfluidics enabled the examination of baseline activities (without external flow) and behaviors in the presence of environmental cues (external flow). A library of qualitative vortex patterns and quantitative hydrodynamics (velocity and vorticity profiles) was generated and shared as a standalone repository. Results from mPIV (velocities) were used to examine the role of ciliary activity in transporting material (oxygen). Given the laminar flow and the viscosity-dominated environments surrounding the larvae, overcoming the diffusive boundary layer is critical for the organism's survival. Péclet number analysis for oxygen transport suggested that ciliary velocities help overcome the diffusion-dominated transport (max Péclet numbers ranged between 30-60). Microfluidics serving as mPIV provided a scalable and accessible approach for investigating the ciliary hydrodynamics of marine organisms. 

\end{sciabstract}

\newpage
\section{Introduction}
The life cycle of most marine invertebrates is divided into a planktonic larval stage (spanning weeks) followed by metamorphosis to bottom-dwelling adulthood (spanning years) \cite{strathmann1971feeding, cowen2009larval, nielsen2018origin}. The free-swimming stage is critical for the success of the individual and the population through seeking new habitats and dispersing \cite{cowen2009larval,gilpin2017vortex}. These behaviors (suspension feeding and swimming) are accomplished by unidirectional beating of cilia located on bands that surround the larva \cite{strathmann1971feeding,strathmann1975larval,gilpin2017vortex}. For feeding in planktotrophic larval forms, there is the additional requirement of using ciliary action to capture unicellular algae for ingestion to fuel the necessary growth and development required for achieving metamorphosis. Planktotrophic larvae ingest algae, but studies indicate that amounts available can fluctuate both spatially and temporally, generally remaining at low levels resulting in food limitations \cite{fenaux1994five, mcalister2018phenotypic,olson1989food,nguyen2021temporal}. These low food concentrations increase planktonic larval duration (PLD), thereby further increasing larval mortality \cite{rumrill1990natural}. Direct investigation of larval feeding and locomotion through the use of ciliary action in the natural environment  is difficult. Therefore, despite their limitations, lab-based strategies have been used to investigate the mechanistic understanding of feeding behaviors. Miniaturized particle imaging velocimetry (mPIV) enabled by microfluidics could provide an effective and scalable strategy to explore the hydrodynamics of suspension-feeding and other ciliary behaviors. Specifically, these systems could control flow conditions to investigate the mechanisms (particularly hydrodynamics) of these behaviors. Beyond their biological importance, knowledge gained through these investigations could advance our understanding of biophysics \cite{gilpin2020multiscale,kanso2021teamwork,nawroth2023flow}, inspiring biomimetic designs for fluidics and robotics \cite{gu2020magnetic,dong2020bioinspired,ren2022soft,zhang2021metachronal}, and provide approaches to mitigating the effects of climate change on marine organisms \cite{levin2006recent,chan2012biomechanics}. 

Among various marine invertebrate larvae, the Pacific sand dollar \emph{(Dendraster excentricus)} is an accessible, abundant, and representative organism (i.e., spheroidal and armed invertebrates) to study the ciliary-mediated behaviors and their hydrodynamics\cite{chan2012biomechanics}. Sand dollars have a feeding pluteus larval stage characterized by an armed morphology \cite{chan2012biomechanics}. Arms are ciliated protrusions that extend from the body and play a role in suspension feeding and swimming behaviors. Sand dollars develop through four, six, and eight-armed stages \cite{clay2010morphology}. The larval stage is followed by the juvenile stage charactrized by the re-absorption of the arms in the rudiment, which becomes the adult sand dollar, and loss of ciliary bands (Figure 1 A and B). Food-induced morphological plasticity is another biologically important aspect of the larval stage \cite{strathmann1971feeding,strathmann1975larval}. Phenotypic plasticity is the ability of a single genotype to produce multiple phenotypes (developmental, physiological, morphological, or behavioral) in response to different environmental conditions \cite{mcalister2018phenotypic}. When food (i.e., unicellular planktonic algae) is limited, sand dollar larvae enhance their feeding ability by increasing the length of their larval arms resulting in longer ciliary bands, which circumscribe the arms. While this response is considered adaptive, it does not fully compensate for the lower food levels, and development is delayed, resulting in a longer PLD. In contrast, when food is abundant, larvae exhibit shorter arm growth and redirect resources to faster growth of postmetamorphic structures (stomach and rudiment), thereby achieving metamorphosis relatively more quickly and shortening their PLD. In their pioneering investigations, Hart and Starthman \cite{hart1994functional}observed that low-fed larvae (long arm phenotype with longer ciliary bands) increased their feeding and clearance rate as measured by particle capture and that plasticity of ciliated band lengths and arm lengths were correlated \cite{hart1991particle, mcalister2018phenotypic}. Recent studies have indicated that morphological plasticity is coupled with significant physiological plasticity \cite{ellison2021different,syverud2023assessing}. In this context, mPIV could be an approach to investigate the relationships between phenotype plasticity, ciliary behaviors, and their hydrodynamics. Beyond this area of focus, sand dollar larvae have been used as a model system to investigate the role of environmental factors (pH, temperature) in the development of marine organisms. Therefore, broader access to simple yet effective tools (i.e., microfluidics) and corresponding analysis strategies could enable high-impact investigations in sand dollar larvae and other marine invertebrates.

The application of microfluidics to study model organisms (e.g., \emph{C. elegans}, zebrafish, and \emph{Drosophila} embryos) has become well-established \cite{hwang2013microfluidic,kim2018microfluidics}. Recently, the use of microfluidics has been extended to other organisms, including \emph{Hydra} \cite{badhiwala2018microfluidics,badhiwala2021multiple,tzouanas2021hydra,hedde2023spim}, \emph{Stentor coeruleus} \cite{blauch2017microfluidic,zhang2021microfluidic,paul2022hydrodynamic},  \emph{Aiptasia larvae} \cite{van2019live}, and planarians \cite{dexter2014chip}. Microfluidics have enabled investigations that would otherwise be difficult (if not impossible), increased experimental throughput, and provided well-controlled physical and chemical microenvironments for housing the organisms. The application of microfluidics to study marine larvae has remained limited. Fluidics and microchambers could provide a valuable approach to examining the hydrodynamics of ciliary-mediated suspension feeding. In a related approach, Gilpin et al. used a 0.5 mm gap between glass slides to examine the ciliary-mediated hydrodynamics of eight-week-old starfish larvae (\emph{Patiria miniata}) \cite{gilpin2017vortex,gilpin2017flowtrace}. Gilpin and co-investigators used beads and microscopy to characterize the underlying ciliary behaviors and the patterns of vortices generated by the starfish larvae. This study provided an actionable framework for future investigations; however, the absence of fluidics prevented the introduction of external flow or other environmental factors. In a related study, Krishnamurthy et al. \cite{krishnamurthy2020scale,krishnamurthy2023active} developed a novel approach for scale-free tracking of suspended organisms (including marine larvae) by combining a ring-shaped circular chamber and a tracking microscope. The approach provided a "hydrodynamic treadmill" to track organisms effectively with no boundaries along the axis of gravity. Inspired by these efforts, this study developed microfluidics with the standard horizontal form factor to investigate ciliary hydrodynamics of the sand dollar larva. The standard form factor of microfluidics introduced boundaries and confinements, which limited the organisms' vertical swimming. However, the standard approach is readily compatible with conventional microscopy and enables the investigation of short-term ciliary hydrodynamics with and without flow.

This effort used microfluidics and fluorescence microscopy as a miniaturized PIV (mPIV) to study ciliary-mediated hydrodynamics of suspension feeding in sand dollar larvae (Figures 1 C and D). Microfluidics were developed using laser-cut molds (4 mm wide and 5 cm long), given the millimeter length scales required to house the organism. First, the feasibility of the hydrodynamic measurements (i.e., fluid flows and vortex fields generated by the larvae) were demonstrated using low and high-fed larvae and fluorescent beads (1.1 $\mu$m in diameter). Vortex patterns were successfully visualized for each larva. Moreover, a juvenile-stage sand dollar was imaged, indicating the absence/significant reduction of the ciliary-mediated vortex features at this stage (Supplemental Figure 2). Two open-source image analysis resources were used to qualitatively (FlowTrace \cite{gilpin2017flowtrace}) and quantitatively (PIVLab \cite{thielicke2021particle}) examine flow patterns. FlowTrace, used as a Fiji plug-in, enabled rapid visualization of flow patterns. PIVLab enabled the characterization of hydrodynamic parameters of the flow, including velocity and vorticity across the field of view and regions of interest. Next, the hydrodynamics of 21 long armed, low-fed larvae across seven days (post-fertilization days 11 to 20) with and without external flow were examined. Sample analyses for 11 and 20 DPF with and without flow (15 $\mu$L.min$^-1$) are provided in the manuscript (Figure 3). The entire library of the analyses (i.e., video recordings, FlowTrace analysis, and the PIV analysis) is shared as part of an Open Science Repository (OSF) page titled: \emph{"Secret Life of Larvae - Hydrodynamics of Dendraster excentricus"}. The same two larvae (11 DPF and 20) were used for additional hydrodynamics, transport, and flow-regime characterizations. Specifically, the average speed for 300 fixed points along a lines of interest (in front of the mouth between the PO arms) were measured (Figure 4). These measurements were valuable since, in addition to PIV videos, it demonstrated the change in the magnitude and direction of bead and fluid velocities. Next, the average velocity for the recording (average velocity vs. time) from the line of interest transport and flow regimes were examined. Velocity values (Vx and Vy) were combined with dimensionless analysis (i.e., 
Péclet and Reynolds number) to examine transport and flow regimes due to ciliary hydrodynamics (Figure 5). Péclet analysis using oxygen as a molecule of interest, suggested that the highest Pé numbers were approximately 60. The analysis suggested that ciliary activity enhanced oxygen transport beyond simple diffusion. The highest Reynolds numbers were less than 1. The Reynolds number analysis suggested that flows generated by the larvae were in the laminar regime. Collectively, these results demonstrated the application of microfluidics as mPIV to assess critical features of ciliary hydrodynamic experiments and their sensitivity to environmental conditions (e.g., flow).

\section{Methods}
\subsection{Larval culture:} Adult sand dollars \emph{Dendraster excentricus} were collected from Los Angeles Harbor in San Pedro, CA, USA ${(33.7088, -118.2806)}$. Larval cultures were conducted using standard protocols as described previously \cite{ellison2021different, syverud2023assessing}. Adult sand dollars were injected with 0.5 M KCl (1 to 2 mL in the coelomic cavity) to induce the contraction of gonadal walls to release eggs or sperm into individual beakers containing sterile filtered seawater.
Concentrated sperm was diluted 1:1000 and mixed with a suspension of eggs to induce fertilization. Fertilization success was confirmed through microscopic observation (Olympus CH-2 Binocular Microscope) of elevated fertilization envelopes approximately 5 minutes after the addition of sperm. 
All cultures were initiated from spawns where fertilization success was at least 90\%. Fertilized eggs were transferred to 20 L vessels at CSULB Marine Laboratory.
Cultures were reared at 16$^{\circ}$C ($^{\pm}$1$^{\circ}$C) and gently mixed using a motor-driven paddle (Buehler Products, Linston, NC, USA) at a speed of 6 rpm High- and low-fed larval feeding treatments were fed 10,000 or 1,000 algal cells mL$^-1$, respectively, of unicellular red algae (\emph{Rhodomonas lens}, strain CCMP739). Algal concentrations were checked and restocked to target concentrations daily using a BD Accuri flow cytometer \cite{rendleman2018more}. Algae cultures were conducted as previously described \cite{syverud2023assessing}, using Erlenmeyer flasks with foam stoppers, and f/2 media. Cultures were harvested at the end of their logarithmic growth phase. Before each feeding, cultures were centrifuged (Beckman Coulter Avanti J-E: 3000 rpm, 12 min, 10 $^{\circ}$C) to remove media.

\subsection{Microifluidic system design and fabrications:} Chips were designed as a straight channel with one inlet and one outlet. Designs are available from the OSF page. The channels were 4 mm wide, 1/16” tall, and 5 cm long. The designs were cut from a 1/16” acrylic sheet (corresponding to the channel height) via a desktop 40 W CO$_{2}$ laser cutter. Laser-cut parts were cleaned with water and attached to plastic or glass slides to form molds. Molds were rested overnight to ensure proper adhesion between the layers before their first use. PDMS (Dow Sylgard™ 184 Silicone 0.5 Kg Elastomer kits) were prepared using the standard protocol, a 10:1 mass ratio of base to curing agent. PDMS was poured on the molds, degassed, and fully cured at 95$^{\circ}$C. PDMS replicas were separated from the mold via a craft knife and tweezers. The inlet and outlet were cut using a 1.5  mm biopsy punch. PDMS replicas were permanently attached to a glass slide via plasma treatment (Harrick Plasma, Basic Plasma Cleaner, maximum RF power 18 W). Devices were used multiple days after their fabrication. Before each use, channel surfaces were rinsed by pipetting seawater (3X 1 mL volumes). Before the experiments, chambers were primed with 16$^{\circ}$C seawater and kept inside the 16$^{\circ}$C incubators. Chambers were reused as part of the experiments. To this end, similar to previous efforts \cite{kim2023phototaxis}, devices were thoroughly cleaned (rinsed to remove beads) and boiled in clean water inside a 1 L beaker on a hotplate (set to 150$^{\circ}$C) for 90-120 minutes. Chambers were discarded typically after 3 to 4 rounds of experiments. 

\subsection{Device loading and experimental procedure:} 
Sand dollar larvae were retrieved from the primary culture (20 L vessels at 16$^{\circ}$C room). A single larva was transferred directly into a previously primed channel for each recording using a partially cut 50-100 $\mu$L pipette tip. Care was taken to release the larvae inside the chip gently. Fluorescent beads (1.1 $\mu$m diameter, Fluoro-Max fluorescent beads) were used to visualize the hydrodynamics and flow pattern. Beads solution were resuspended in 1:100 ratio in 0.2 $\mu$m filtered seawater. Solutions were vortexed before each experiment. For the experiments with flow, constant infusion (at 15 $\mu$L.min$^-1$ of 16$^{\circ}$C seawater containing beads) was used. A syringe pump (KDS Legato) was used for constant infusion. A visualization of the material used in the experiments is provided in Supplementary Figures (Figure S1). Microfluidics were imaged using a fluorescence microscope. The microscope was controlled via Micromanager software \cite{edelstein2014advanced}. 

\subsection{Image and video analysis:} FlowTrace was used as an ImageJ/Fiji plugin, as described previously, to provide a rapid qualitative visualization of flow patterns \cite{gilpin2017flowtrace}. Two dimensional PIV analysis were used to quantify the flow. The analysis used PIVLab software\cite{thielicke2021particle}, image were analyzed using a shifting overlaping windows (decreasing sizes of 64 × 64 pixels, then 32 × 32 pixels, then 16 x 16 pixels). Additional resources and instructions for PIVLab are shared in the project repository.

\subsection{Dimensionless numbers analysis:} Two dimensionless numbers Péclet and Reynolds numbers were used to examine the transport due to ciliary behaviors of larvae. 

\[ Equation-1: Pe = \frac{RU}{D}\]

R: Larvae radius (estimated as a circle - see Figure 4)\\
U: The maximum Vx or Vy for each frame from line of interest (LOI)\\
D: Diffusion coefficient of oxygen in water ($10^{-9} m^{2}.s^{-1}$)\\

\[Equation-2: Re = \frac{\rho UR}{\mu}\]

$\rho$: Fluid density 1000 $kg.m^{-3}$ at 20$^{\circ}$C \\
U:  The maximum Vx or Vy for each frame from LOI\\
R:  Larvae radius (estimated as a circle)\\
$\mu$:  Dynamic viscosity $10^{-3} Pa.s$ at 20$^{\circ}$C \\

\section{Results}
\subsection{Low- and High-Fed Larvae:} 
Vortex hydrodynamic behaviors of low-  and high-fed larvae and a juvenile sand dollar were visualized to demonstrate the approach's feasibility. Each organism was transferred into the microfluidics with 1.1 $\mu$m diameter beads to obtain recordings. Recall that low-fed organisms have longer arms and ciliary bands than high-fed organisms. The FlowTrace algorithm was used to visualize the hydrodynamic vortex patterns generated by each larva (Figure 2. A and B, and Supplementary Videos). Two symmetric (across the main body-axis) vortex fields were visible. Next, using PIVLab, the ciliary behaviors and corresponding fluid velocity across the field were quantified. These experiments suggested that higher bead concentrations were more suitable for the PIV analysis (please note that the low-fed recording had fewer beads than the high-fed recording). Next, the hydrodynamics were examined for juvenile sand dollars. Sand dollar’s life cycle includes a bilaterally symmetric larval stage that metamorphoses into a benthic (non-swimming) pentaradially symmetric juvenile stage, which grows into the adult sand dollar. In this juvenile stage, ciliated larval arms are absent, and the spherical body is covered with calcium carbonate spines and hydraulic tube feet. The recordings and corresponding analysis indicated the absence (or reduction) of  vortex flow patterns in the juvenile stage (Figure S2). These experiments established the feasibility of the experiments and corresponding protocols.

\subsection{Tracking hydrodynamics of low-fed larvae:} 
Using low-fed larvae (n=21 larvae), hydrodynamics from 11 days post-fertilization (DPF) to 20 DPF were monitored. To this end, recordings with (15 $\mu$L.min$^-1$) and without flow for 3 larvae at each developmental point were obtained. Care was taken to obtain recordings when the larvae were in a vertical position (Anterior view) and during suspension-feeding without swimming (See Figure 1 D). Hydrodynamics for each recording (velocity and vorticity fields) were obtained using PIVLab. Sample analysis for two larvae from the start (11 DPF) and end of the culture (20 DPF) are presented (Figure 4). After each recording, larvae were retrieved from the chips and imaged using a binocular microscope (Olympus CH-2 Binocular Microscope) to quantify morphological parameters (as defined in Figure 1 E). These parameters included post oral (PO) arm length, the angle between the body and the PO arms, body length, and the distance between the two PO arm tips (used in this study). The summary of morphological parameters is provided as supplemental material (Figure S3). The corresponding repository contains all the recordings (i.e., video recordings and FlowTrace) and PIV analysis (i.e., vorticity and velocity fields). 

\subsection{Velocity and transport characterizations via dimensionless numbers:} 
Next, the hydrodynamics and transport due to ciliary activity were measured. Two larvae from the start  (11 DPF) and end of the culture (20 DPF) were selected. The spatial dynamics of fluid velocities were first examined using a line of interest (LOI) (see Figure 4 dashed lines). This analysis measured the average velocities (Vx and Vy components) for 300 fixed points across the LOI for experiments with and without flow (Figure 4). Analysis (similar to the PIV videos) indicated the opposite directions of the flow (for example, note Vy components). PIV results (velocities) were then used to examine oxygen transport. To this end, each frame's average velocity (across the entire line of interest) was calculated (Figure 5). These velocities were used in a dimensionless analysis, similar to the efforts by Kanso et al. \cite{kanso2021teamwork}. Larvae were approximated as circles (Figure 4 dashed circles around the larvae). To estimate the Péclet number (Equation 1), the maximum values of Vx and Vy for each frame of the recording from the LOI were used. The highest Pé numbers (due to velocities in either direction) were between 30 to 65. Recall when Péclet less than one, molecular diffusion dominates mass transport. For the corresponding velocities, the range of the Pé number suggests that the transport is advection-dominated. The analysis suggested that ciliary hydrodynamics improved oxygen transport beyond diffusion alone. Moreover, each larvae’s Reynolds number (Equation 2) was calculated. Typically, the highest Re numbers were <0.06, suggesting laminar flow regimes. 

To further examine the transport phenomena, PIV recordings for no-flow conditions were used to estimate the aggregate bead speeds generated by each larva. To do so, a region of interest (ROI), informed by the size of the larvae (i.e., each side is twice the length of distance between post oral arms of the larva being analyzed), was established for each recording. Using PIVLab, the net average speed (instead of velocity) inside the ROI for each recording frame was calculated (Figure S4). Each larva's speed values were used to estimate an average Péclet number using oxygen as the molecule of interest (see Figure S5). The corresponding Pé numbers for all larvae ranged between 10 and 30 (see Supplemental Table-1). Suggesting similar results between the ROI and the spatially informed LOI analysis.   

\section{Discussions}

In the work presented here, we developed simple microfluidics as mPIV to investigate suspension-feeding hydrodynamics in sand dollar larvae. Channels widths of 4 mm and 1.5 mm tall were used to provide sufficient space for the suspension feeding behaviors of the larvae. The standard form factor of the chip made it possible to readily interface with conventional microscopes. However, this form factor limited the vertical swimming of the larvae. 
For future experiments, the orientation of a chip (vertical instead of a standard horizontal form factor) can be modified to enable these behaviors. To this end, alternative imaging approaches similar to side imaging efforts by Krishnamurthy et al. and Le et al. \cite{krishnamurthy2020scale,le2023orthogonal} may be required.
Recordings were obtained when the larvae were upright (anterior view) and not swimming to consider suspension feeding behaviors. In this investigation, 1.1 $\mu$m diameter, microbeads were used since the beads had a similar diameter to the unicellular algae (\emph{Rhodomonas lens}, strain CCMP739) that is fed to the larvae in the laboratory setting. Plastic beads have been previously used to study larval feeding mechanisms \cite{hart1994functional}. In future experiments, algae or micro-particles with varying diameters or mechanical properties could also be considered \cite{pernet2018larval}.

In the first experiment, vortex hydrodynamics of low- and high-fed larvae were investigated using qualitative (FlowTrace) and quantitative (PIVLab) approaches. A juvenile sand dollar was also imaged, demonstrating the absent hydrodynamic patterns. These experiments established the feasibility of the mPIV approach. While the relationships between hydrodynamics and phenotypic plasticity are of great interest, current effort focused on tracking low-fed larvae's suspension feeding. To this end, we sought to track the hydrodynamics and transport of low-fed larvae across multiple days. The microfluidic system, in particular, was used as an effective mPIV to quantify the flow velocities generated by the larvae. In the subsequent experiments, the hydrodynamics of 21 (7 days, 3 larvae each day) low-fed larvae were tracked from 11 DPF to 20. Here, ciliary-hydrodynamics with and without external flow were examined for each larva. The external flow rate was kept constant at 15 $\mu$L.min$^-1$. This value was selected since the larvae could tolerate it without getting washed away.
Future experiments will focus on the roles of velocity profiles and shear forces during larval development in morphologically divergent low- and high-fed larvae. Additionally, in a microfluidic system, the role of chemical cues or mechanical forces can be examined by taking advantage of the laminar flow profiles inside a chip. The complete library of quantitative (PIV) and qualitative (FlowTrace) analysis for each larva has been shared with a standalone repository. To carefully examine the flow profiles, two larvae, one from the start of the campaign (11 DPF) and one from the end (20 DPF), were selected for additional characterizations. First, a LOI was used to obtain spatially informed velocities. These values were used to estimate the  Pé numbers for oxygen transport. The range of maximum Pé numbers (30 to 60) suggested that the transport is advection-dominated. It is well-established that small microbes (cells <1$\mu$m in diameter) can secure nutrients and metabolites via diffusion alone. Since the primary use of diffusion would lead to creating a diffusive boundary layer (DBL) and the depletion of resources needed for supporting the organism \cite{kaiser2011marine}. Therefore, many marine organisms use strategies to overcome DBL allowing them greater access to surrounding food and dissolved nutrients. For example, swimming, sinking, forming colonies, and even symbiotic relationships \cite{kanso2021teamwork} have been observed to help overcome DBL. However, many questions remain about the relationships between environmental factors and organisms' size and behaviors. For example, it has been suggested that swimming is primarily helpful for organisms with a diameter larger than 100 $\mu$m; in contrast, organisms between 1 to 100 $\mu$m diameter use swimming to move to regions with higher nutrient concentration \cite{kaiser2011marine}. Here, microfluidic systems could be used to carefully examine these relationships between organism morphology, behaviors, and their ability to overcome DBL across different scenarios (flow regimes or food concentrations). Next, using a variation of PIV analysis (ROI instead of LOI), each larvae's average speed (across the entire recordings) was calculated. This analysis did not include the first analysis's spatial (Vx and Vy) component. However, it provided a comprehensive assessment of the change in speed around (as opposed to in front) the larvae. These values were used to estimate Pé numbers for oxygen. Again, the range of values suggested advective transport enabled via ciliary flows. Collectively, these analyses suggested the feasibility of obtaining hydrodynamics and transport from the approach. It is important to note that future investigations should monitor more individuals (currently n=3) for each DPF and start the campaign as early as possible. This study provides the methods (microfluidics, imaging, and analysis) that would enable future investigations.  

\section{Conclusions}

Microfluidic systems have enabled investigations of model organisms, including \emph{C. elegans}, zebrafish, and \emph{Drosophila} embryos  \cite{hwang2013microfluidic,kim2018microfluidics}. More recently, microfluidics has been successfully applied to study other organisms such as \emph{Hydra}, \emph{Stentor coeruleus}, \emph{Aiptasia larvae}, and planarians \cite{badhiwala2018microfluidics,blauch2017microfluidic, van2019live, dexter2014chip}. Similar to these efforts, microfluidics could provide scalable strategies to investigate key aspects of marine larvae hydrodynamics, development, and underlying biology. For example, the related roles of feeding and swimming behaviors are of interest. It has been suggested that investigating "how larvae feed" and their mechanistic models are fundamental questions across various lines of inquiry in the field \cite{pernet2018larval}. For example, the relationships between clearance rate and the particle size range that larvae can catch are unknown \cite{pernet2018larval}. Direct observation of these behaviors and their hydrodynamics, as opposed to the average quantification of the decline rate for the particles over time, could provide a mechanistic description of feeding behaviors, their underlying hydrodynamics, and transport. Microfluidics could provide a versatile and scaleable approach for these experiments. Moreover, these approaches can later connect the feeding behaviors to the questions related to phenotypic plasticity (both morphological and physiological).

Marine invertebrates exhibit dynamic and complex hydrodynamics during their planktonic larval stage. In the case of sand dollar larvae, the hydrodynamics are primarily due to the ciliary bands that cover the organism's body and arms. Advancing our knowledge of ciliary behaviors (across different organisms) is of great interest in biology and biomimetic design \cite{nawroth2023flow,gilpin2020multiscale,bull2021ciliary}. To this end, approaches to broaden access to examine the ciliary behaviors and their underlying hydrodynamics (e.g., vortex patterns and transport) are needed. In one example, the approach could broaden our understanding of the interactions between hydrodynamics, physical cues, and underlying mechanisms (e.g., mechanosensing via cilia). Microfluidics could serve as a versatile technique to modulate the microenvironment of larvae across various physicochemical factors (e.g., flow, temperature, pH, salinity, viscosity, and presence of chemicals). Microfluidics could provide a scalable approach to investigate the interaction between the larvae (using hydrodynamics or alternative outputs) to investigate the emerging environmental threats due to climate change or microplastic pollution. Similar to all experimental techniques, microfluidics have limitations. For example, given the chip boundaries, attention has to be paid to the role of confinement and how they may inform patterns of behaviors. Specifically, the system cannot be readily used for vertical swimming in its current form. However, with modifications, many of these challenges can be overcome. From an engineering perspective, investigating the hydrodynamics of marine larvae could provide valuable metrics and designs for the biomimetics design of next-generation robotics and control elements for fluidics. Therefore, broader access to the tools (next-generation fluidics, novel imaging, computational and modeling systems) and future collaborations are needed. 

\newpage
\section* {Acknowledgments}  
Authors express gratitude to members of Ahrar-lab and Pace-lab.\\
Authors are thankful to CSULB Marine Lab, specifically Yvette Ralph.\\
Authors acknowledge Samuel Korban for the life-cycle illustration developed for Figure 1.B.\\
Authors express gratitude to Prof. Albert Siryaporn and Prof. Rob Steele for their feedback on the manuscript.\\

This work was supported in part by CSULB startup funds, a CSUPERB new investigator grant to S. Ahrar, CSULB COAST student awards, and a STEM-NET collaborative seed award to S. Ahrar and D. Pace.\\

W. Chen was supported by a CSULB NIH-BUILD (NIH; UL1GM118979; TL4GM118980; RL5GM118978) scholarship.\\

\section*{Corresponding Authors Contacts}

Siavash Ahrar (Ph.D.)\\
Mail: Department of Biomedical Engineering, California State University Long Beach, Long Beach, CA 90840, United States of America.\\   
E-mail: Siavash.Ahrar@csulb.edu \\

Douglas A. Pace (Ph.D.)\\
Mail: Department of Biological Sciences, California State University Long Beach, Long Beach, CA 90840, United States of America.\\
E-mail: Douglas.Pace@csulb.edu\\

\section*{Affiliations}

\begin{itemize}
\item 1: Department of Biomedical Engineering. California State University Long Beach, CA, US.

\item 2: Department of Biological Sciences. California State University Long Beach, CA, US.
\end{itemize}

\section* {OSF Repository}  

OSF Name: Secret Life of Larvae - Hydrodynamics of \emph{Dendraster excentricus}/
Link: \url{https://osf.io/fqwkb/}

As part of the project, resources, including the complete library of videos, design files, and code for analysis, have been shared.  

\section* {Conflict of interest}  
Authors declare no conflict of interest. 

\section* {Institutional Review Board Statement}  
Not applicable. 

\newpage
\textbf {List of Figures}:
\begin{itemize}
  
  \item Figure 1: Hydrodynamic investigation of the Pacific sand dollar larvae \emph{Dendraster excentricus}.
  \item Figure 2: Hydrodynamic measurements from low- and high-fed larvae.
  \item Figure 3: Sample PIV measurements from low-fed larvae from 11 and 20 DPF with and without flow.
  \item Figure 4: Average velocity for 300 fixed points across the line of interest
  \item Figure 5: Transport and flow-regime characterized due to ciliary  hydrodynamics.
\end{itemize}

\textbf {List of Supplementary Figures}:
\begin{itemize}
  
  \item Sup Figure 1: Visual bill of materials and supplies. 
  \item Sup Figure 2: Hydrodynamic measurements of a juvenile sand dollar.
  \item Sup Figure 3: Morphological parameters.
\item Sup Figure 4: Changes in average speed in time due to ciliary behaviors inside the ROI (DPF 11 to DPF 20).
\item Sup Figure 5: Péclet numbers based on average speed.
\item Sup Figure 6: Relationship between Péclet number and larval properties.

\end{itemize}

\textbf {List of Supplementary Videos}:
\begin{itemize}
  
  \item Movie S1: Hydrodynamics of low-fed larvae
  \item Movie S2: Hydrodynamics of high-fed larvae
  \item Movie S3: Hydrodynamics of a juvenile sand dollar
  \item Movie S4: PIV measurements from low-fed larvae from 11 DP with no flow
  \item Movie S5: PIV measurements from low-fed larvae from 11 DP with flow
  
\end{itemize}

\newpage
\begin{flushleft}
\bibliographystyle{unsrt}
\bibliography{refs} 
\end{flushleft}

\newpage

\begin{figure}[b]
\includegraphics[width=\textwidth]{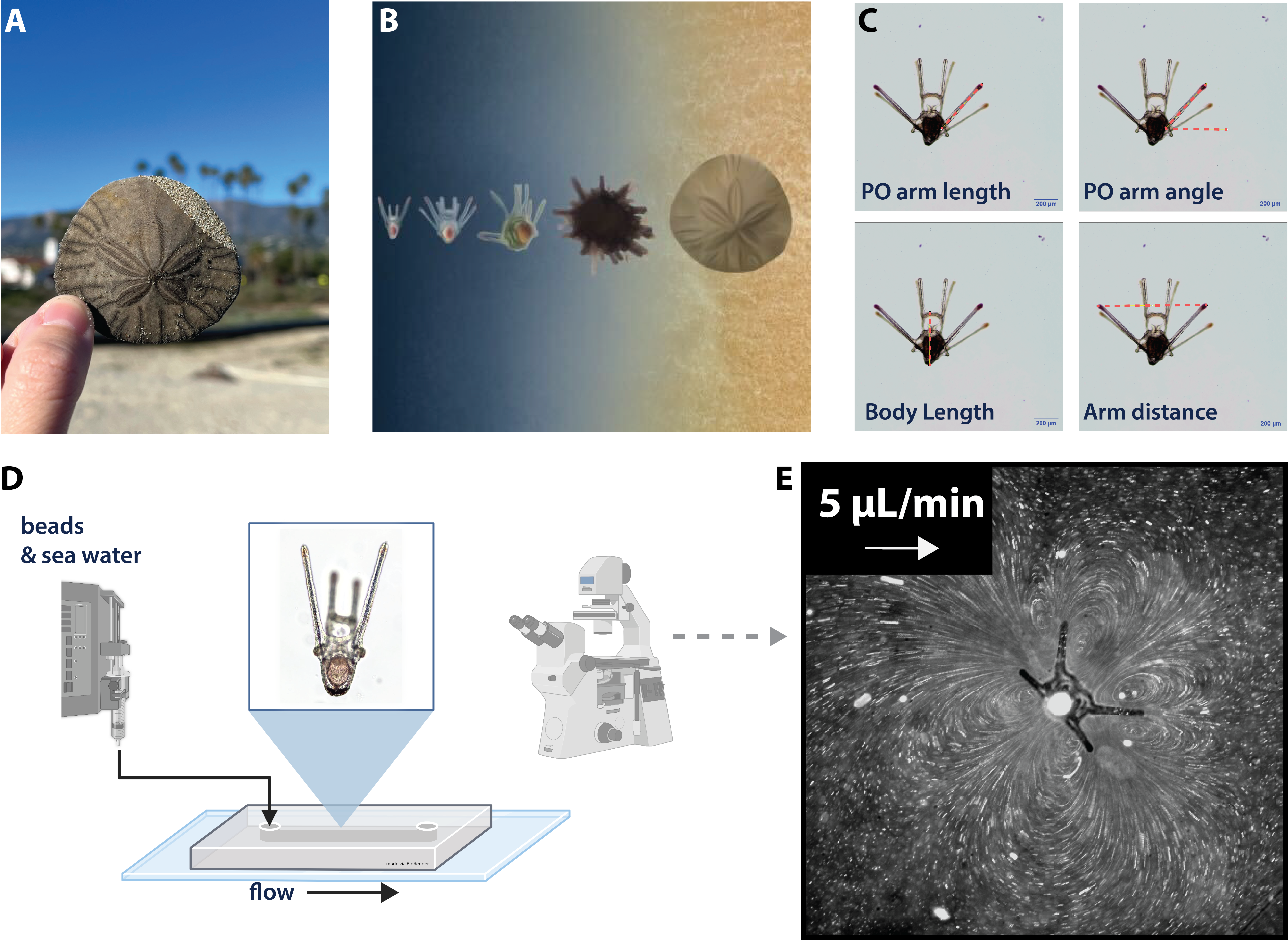}
\centering
\caption{\textbf {Hydrodynamic investigation of the Pacific sand dollar larvae (\emph{Dendraster excentricus}).} (A) An adult Pacific sand dollar from the coast of California (Santa Barbara, California, United States). Sand dollar larvae are an abundant and representative model for the hydrodynamic investigation of marine organisms. (B) Illustration of larval growth (first three diagrams), juvenile stage, and bottom-dwelling adulthood. The free-swimming larval stage is the primary stage where large-scale dispersal is possible and is therefore a critical stage for recruitment to adult populations.  (C)  Examples of well-established morphological parameters from the larvae include the post oral (PO) arm, PO arm angle, and body length. This study also measured the distance between the arm lengths for each larva (ventral view).(D) Microfluidics and microscopy are combined as a miniaturized PIV (mPIV) apparatus to investigate the hydrodynamics of the larval stage (anterior view). (E) Ciliary-mediated hydrodynamics (vortex patterns) generated by larvae were visualized using 1.1 $\mu$m diameter beads. Microfluidics can be used to include flow in well-controlled environments. }
\end{figure}

\begin{figure}[b]
\includegraphics[width=\textwidth]{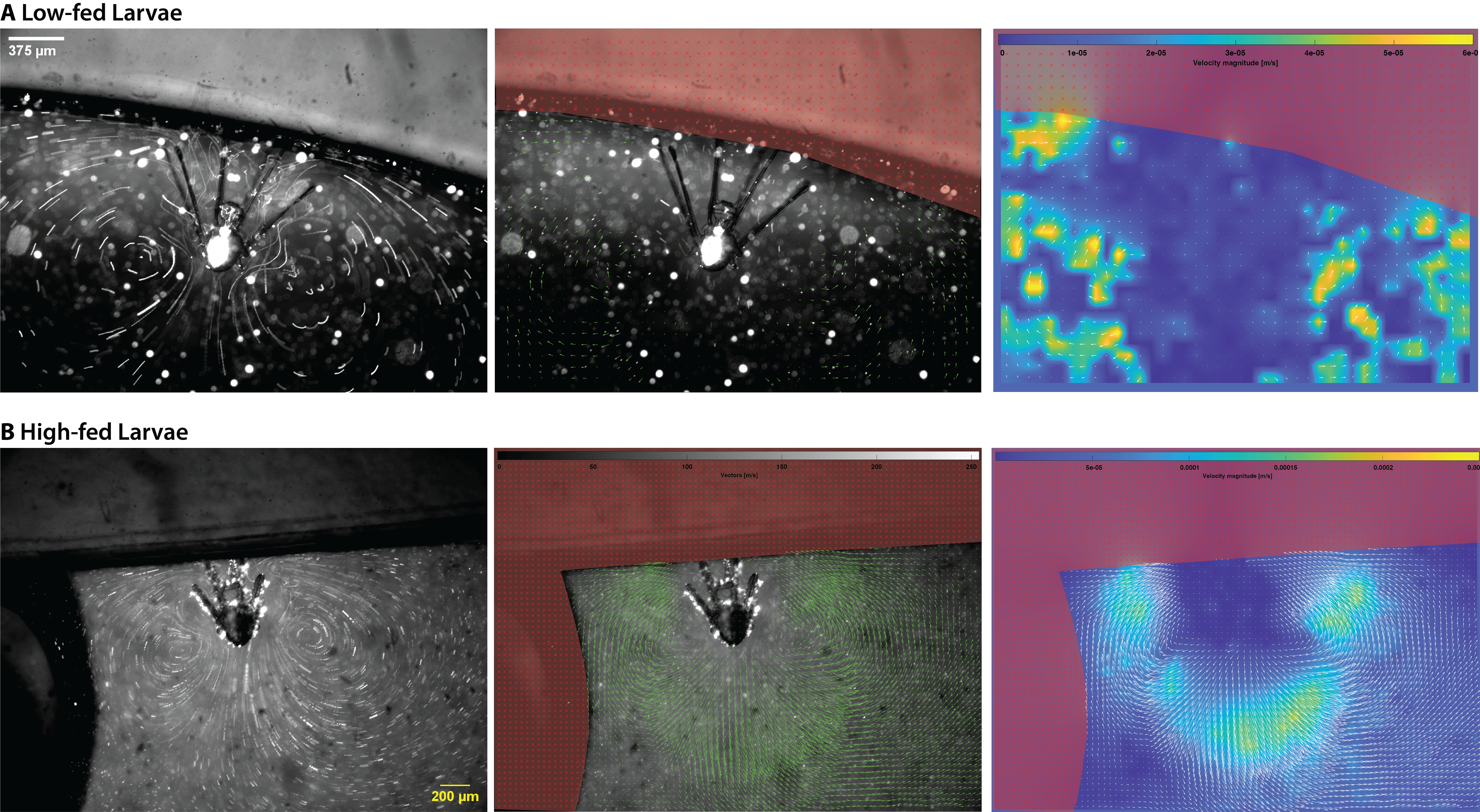}
\centering
\caption{\textbf {Hydrodynamic measurements from low- and high-fed larvae.} (A) Low-fed larvae have longer arms, while high-fed larvae grow faster. Symmetric vortex patterns were visualized across the body-axis  (ventral view). The qualitative analysis indicates the presence of ciliary-mediated vortex patterns.  PIV-enabled velocity vectors  (directions) and PIV-enabled velocity vectors (magnitude) enables the quantification of these behaviors in space and time. 
(B) Cillary hydrodynamics of high-fed larvae. Similarly flow features can be observed and quantified inside the microfluidics. Experiments established the feasibility of hydrodynamic measurements with microfluidics. Experiments also suggested that higher bead concentrations were better suited for PIV analysis. Note the difference between the low-fed with lower bead concentrations and the high-fed with higher bead concentrations.}
\end{figure}

\begin{figure}[b]
\includegraphics[width=\textwidth]{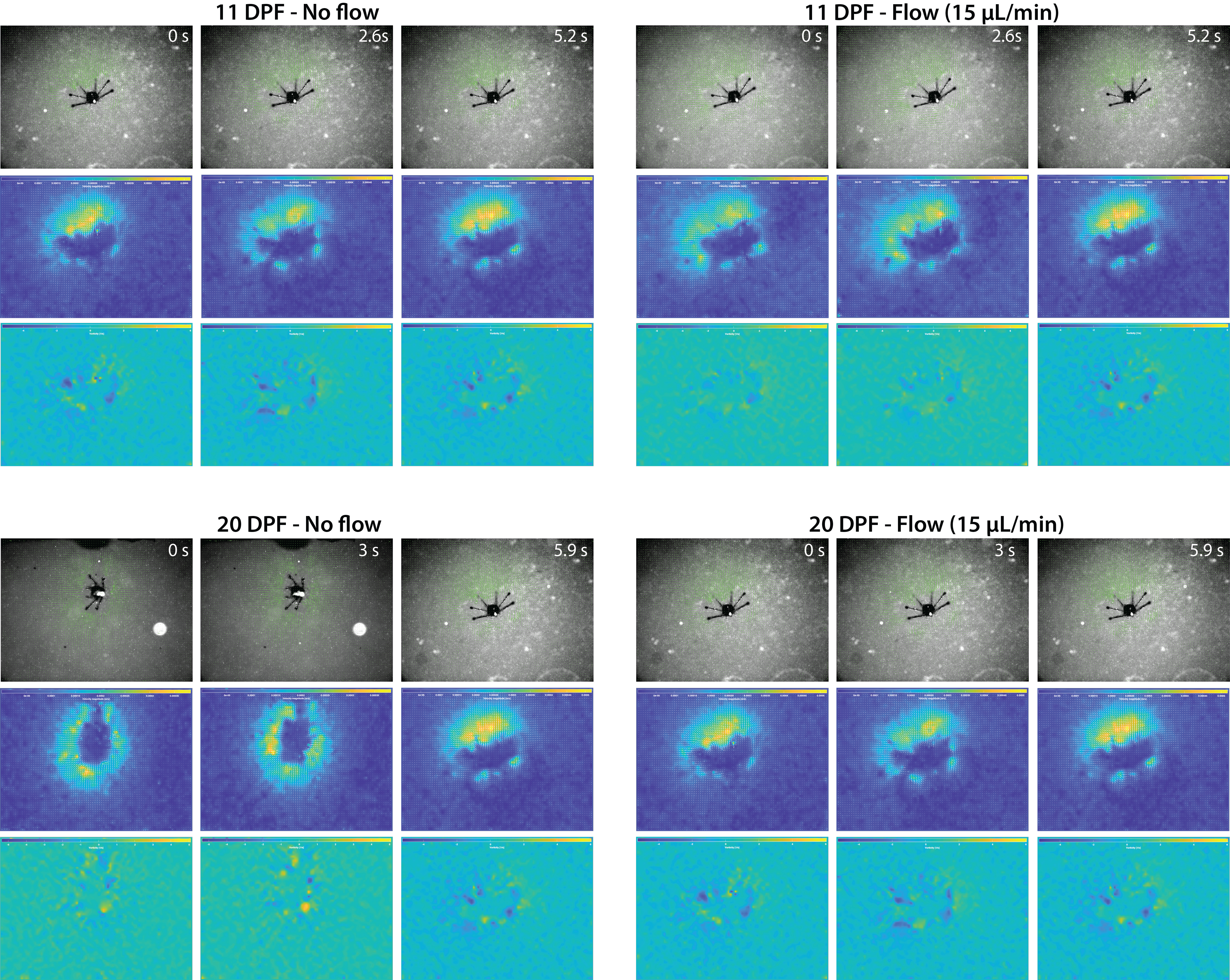}
\centering
\caption{\textbf {Sample PIV measurements from low-fed larvae from 11 and 20 DPF with and without flow.}  PIV measurements include the velocity vectors, which emphasize the direction and magnitude of the flow; velocity heat maps, which emphasize the magnitude of the flow; and vorticity vectors, which emphasize the direction and magnitude of fluid and beads rotation/spinning due to the ciliary hydrodynamics of the larvae.}
\end{figure}

\begin{figure}[b]
\includegraphics[width=\textwidth]{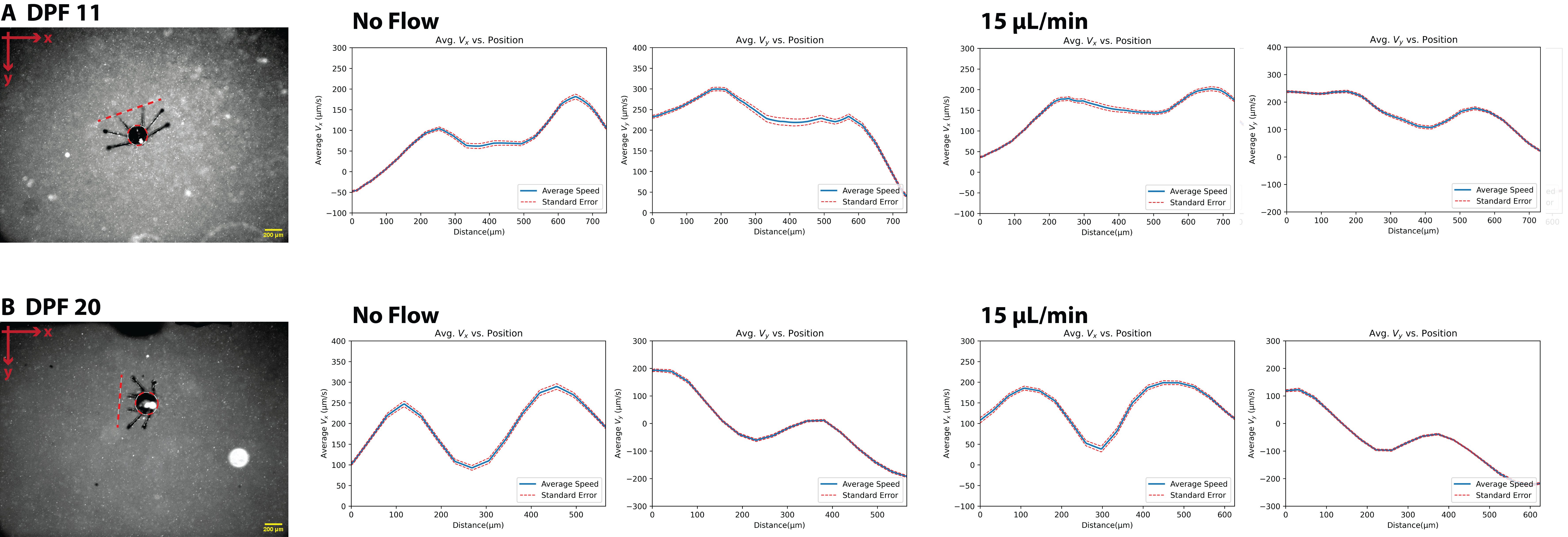}
\centering
\caption{\textbf {Average velocity for 300 fixed points across the line of interest.} (A) Bead velocities at 11 DPF with and without flow. Line of interest (LOI) was defined based on the distance between two PO arms. 
(B) Bead velocities at 20 DPF variations in velocity with and without flow.}
\end{figure}

\begin{figure}[b]
\includegraphics[width=\textwidth]{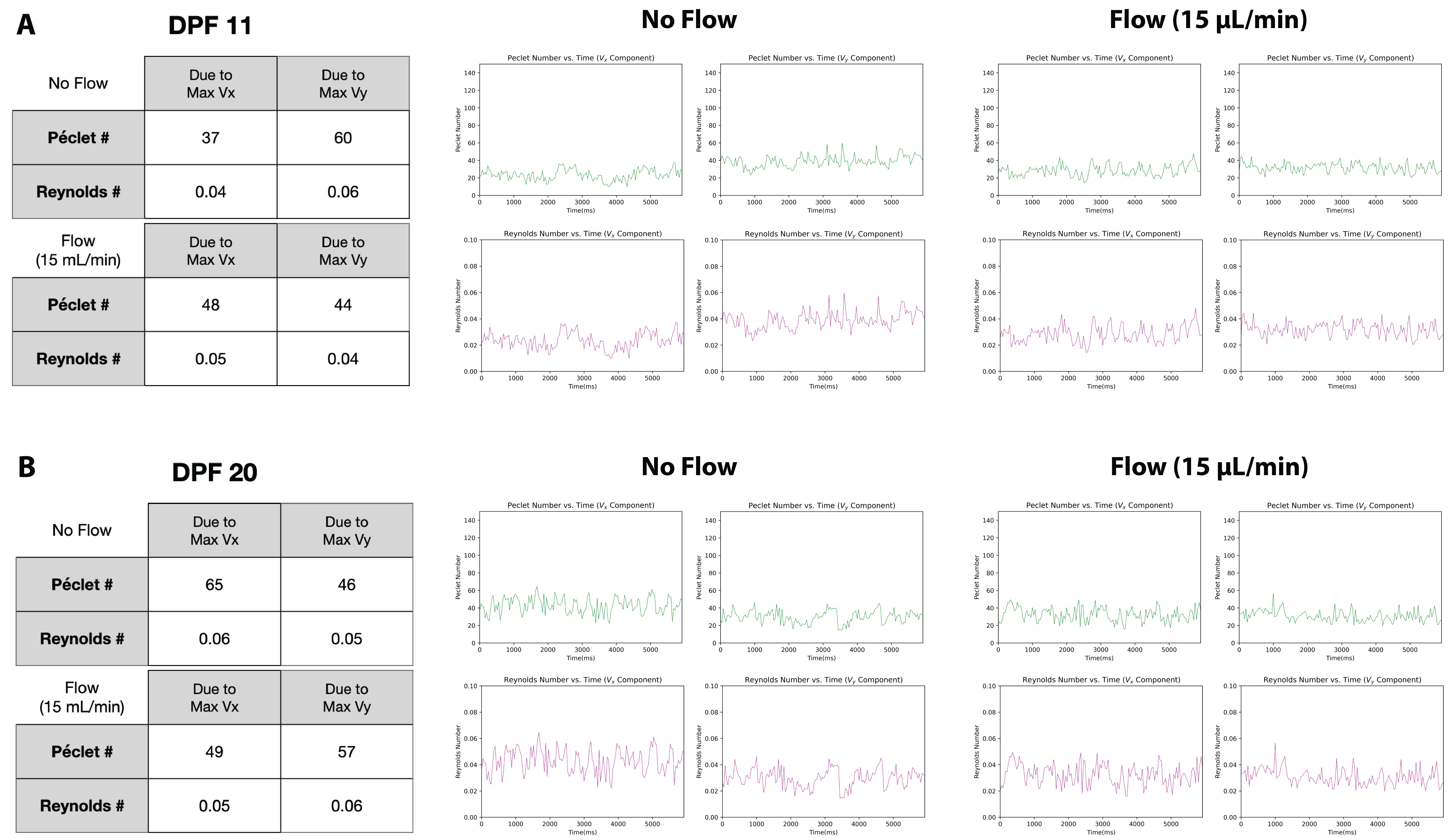}
\centering
\caption{\textbf {Transport and flow regime characterized due to ciliary-mediated hydrodynamics.} (A) Results from the larvae at 11 DPF, the table summarizes the highest Péclet number (diffusion coefficient of oxygen was used) and Reynolds number due to maximum values of Vx and Vy. Plots demonstrate the change in the maximum  Péclet and Reynolds numbers. Maximum Pé was between 40 and 60, which suggested the ciliary flows enhanced the active transport of oxygen. The Reynolds number analysis suggested that the flows were in laminar regimes. (B) Results from the larvae at DPF, the maximum Pé ranged from 45 to 65, and the maximum Re ranged from 0.05 to 0.06. Analysis suggest that the transport properties are stable during this stage of growth.}
\end{figure}

\begin{figure}[b]
\includegraphics[scale=0.15]{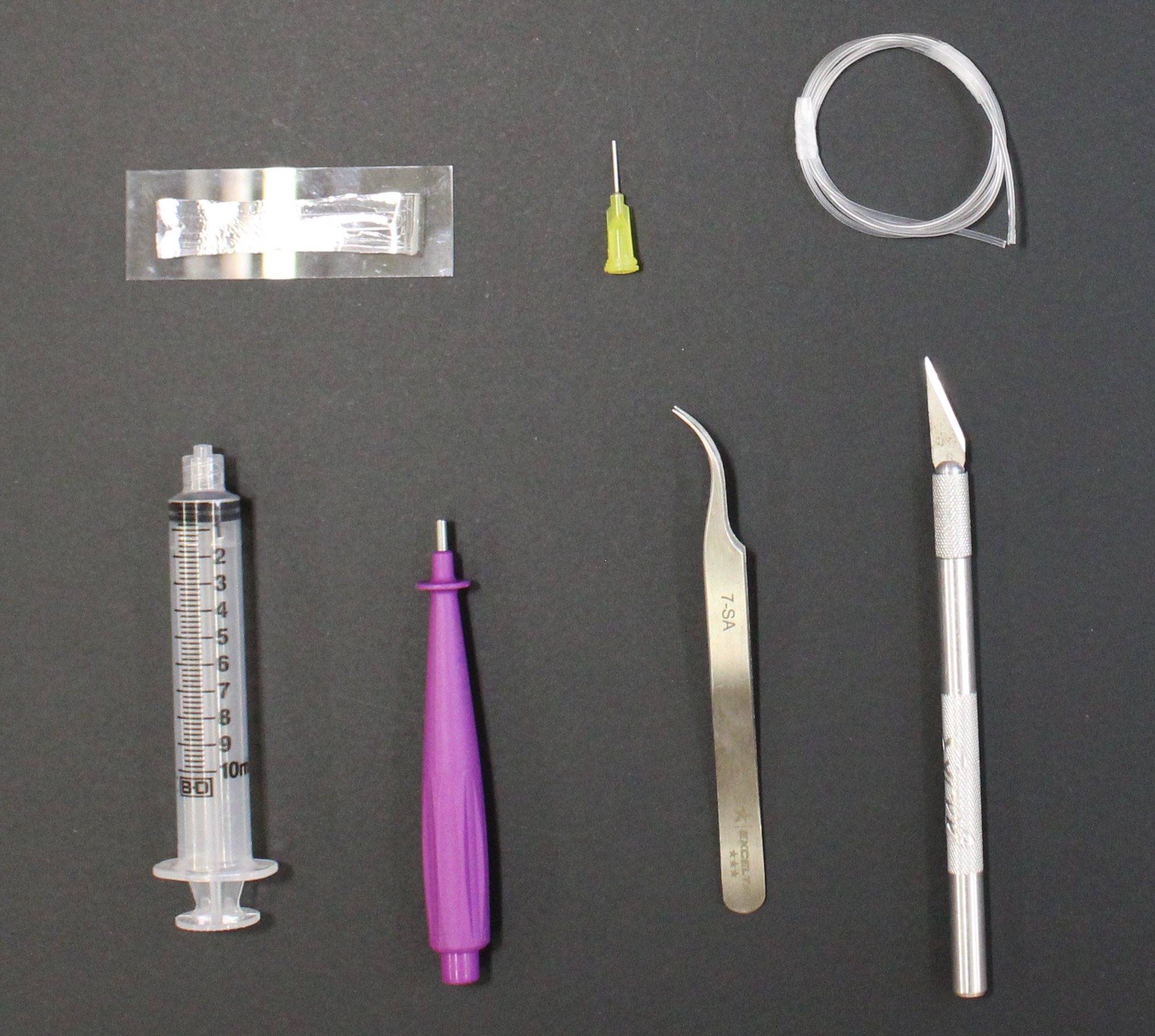}
\centering
\caption{\textbf {Supplemental Figure 1 - Visual bill of materials.} A complete list of materials is provided in the repository.}
\end{figure}

\begin{figure}[b]
\includegraphics[scale=0.75]{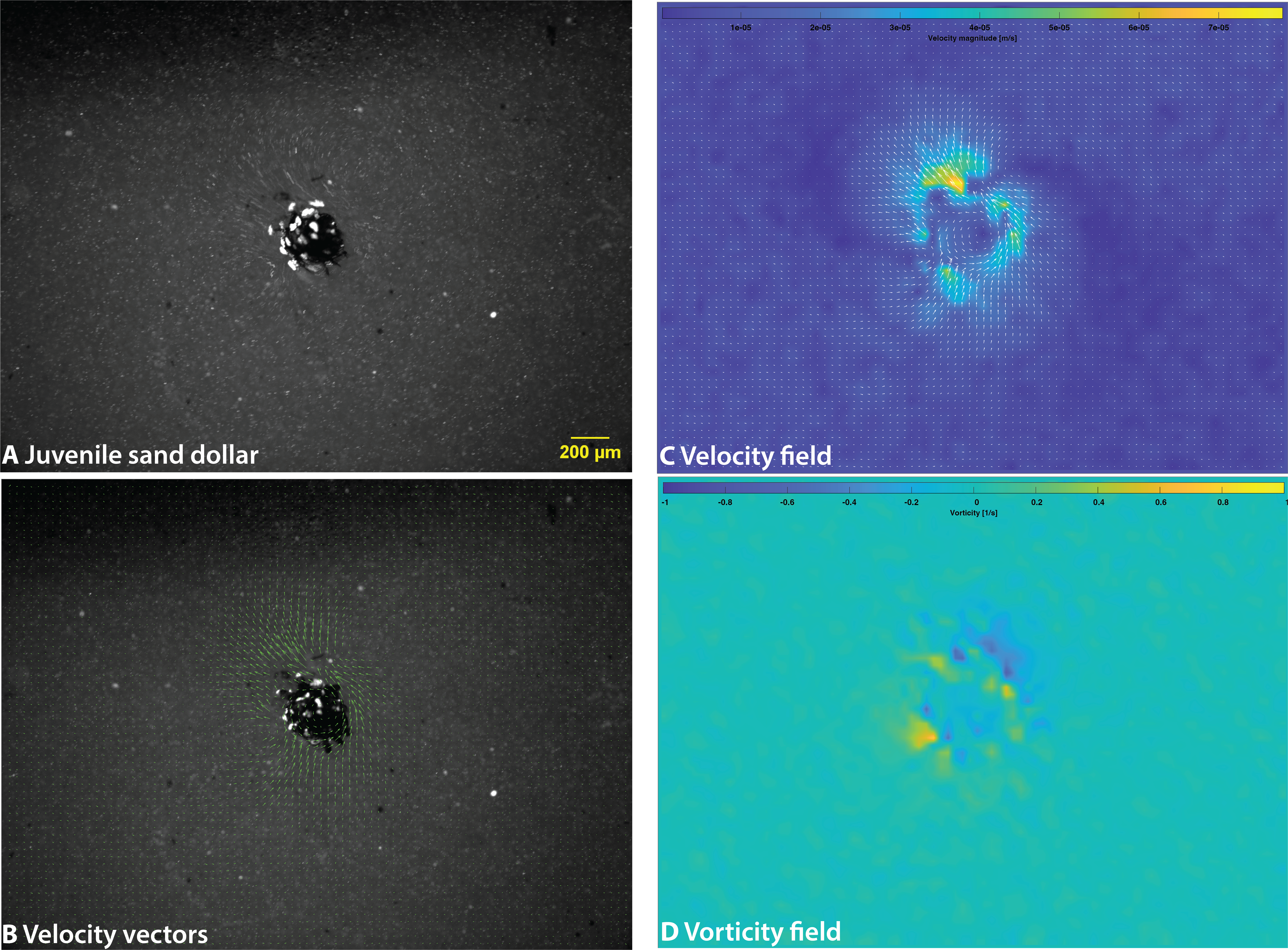}
\centering
\caption{\textbf {Supplemental Figure 2 - Hydrodynamic measurements of a juvenile sand dollar.} (A) A qualitative analysis indicates the absence or reduction of ciliary-mediated vortex patterns. (B) PIV-enabled velocity vectors  (directions)  (C) PIV-enabled velocity vectors (magnitude), and (D) Vorticity across the entire field of view.}
\end{figure}

\begin{figure}[b]
\includegraphics[scale=0.75]{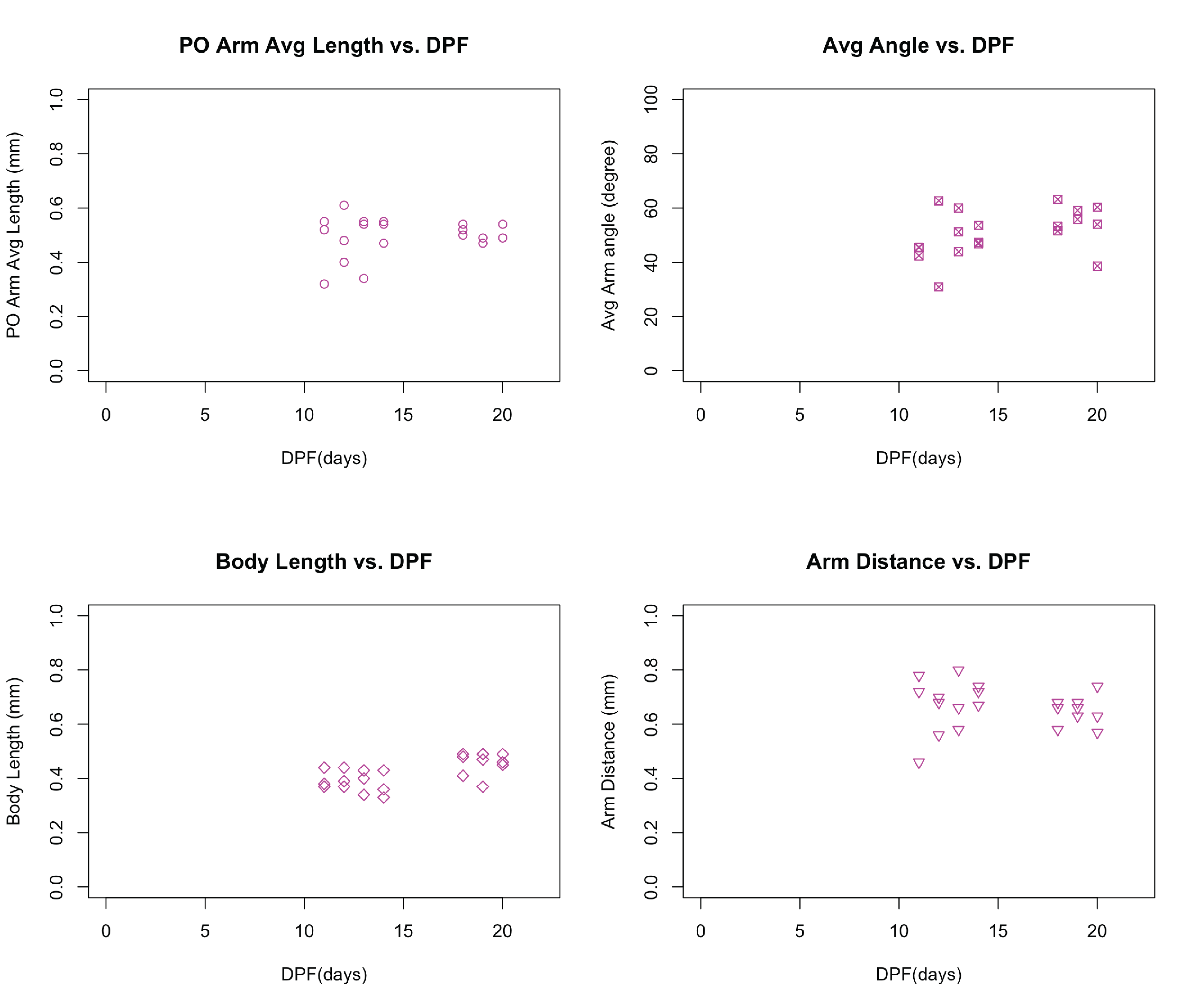}
\centering
\caption{\textbf {Supplemental Figure 3 - Morphological parameters.} Post oral (PO) arm length, PO arm angle, and Body length are well-established morphological parameters. The arm distance was also measured in this study. 
}
\end{figure}

\begin{figure}[b]
\includegraphics[width=\textwidth]{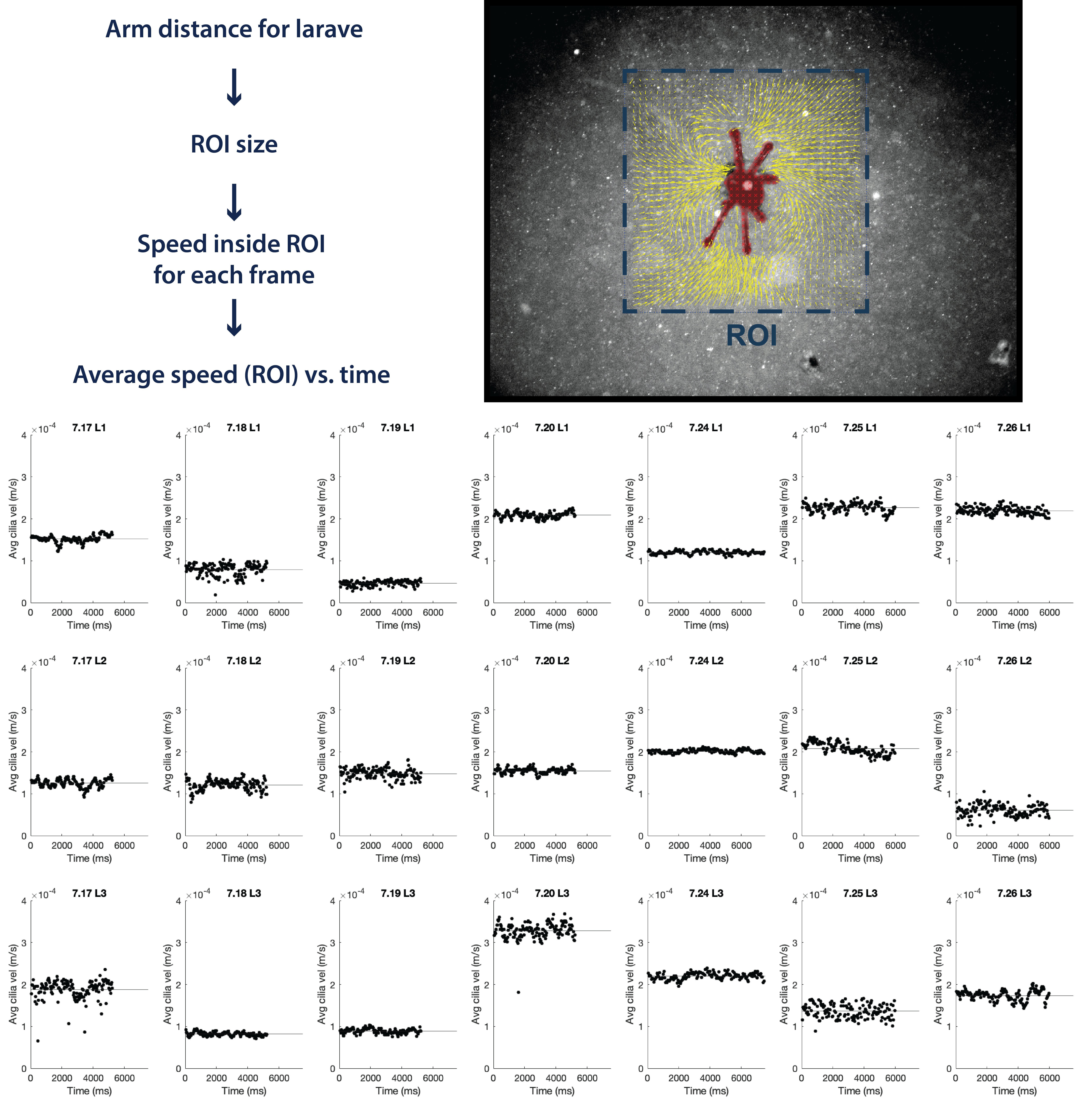}
\centering
\caption{\textbf {Supplemental Figure 4 - Changes in average speed of the ciliary behaviors inside the region of interest (ROI) for each 21 larvae from DPF 11 to DPF 20.} A ROI for each larva was defined based on each larvae’s arm distance. The size of ROI was 2x the size of PO arm length.  The average speed for each frame inside the ROI were measured via PIV.}
\end{figure}

\begin{figure}[b]
\includegraphics[width=\textwidth]{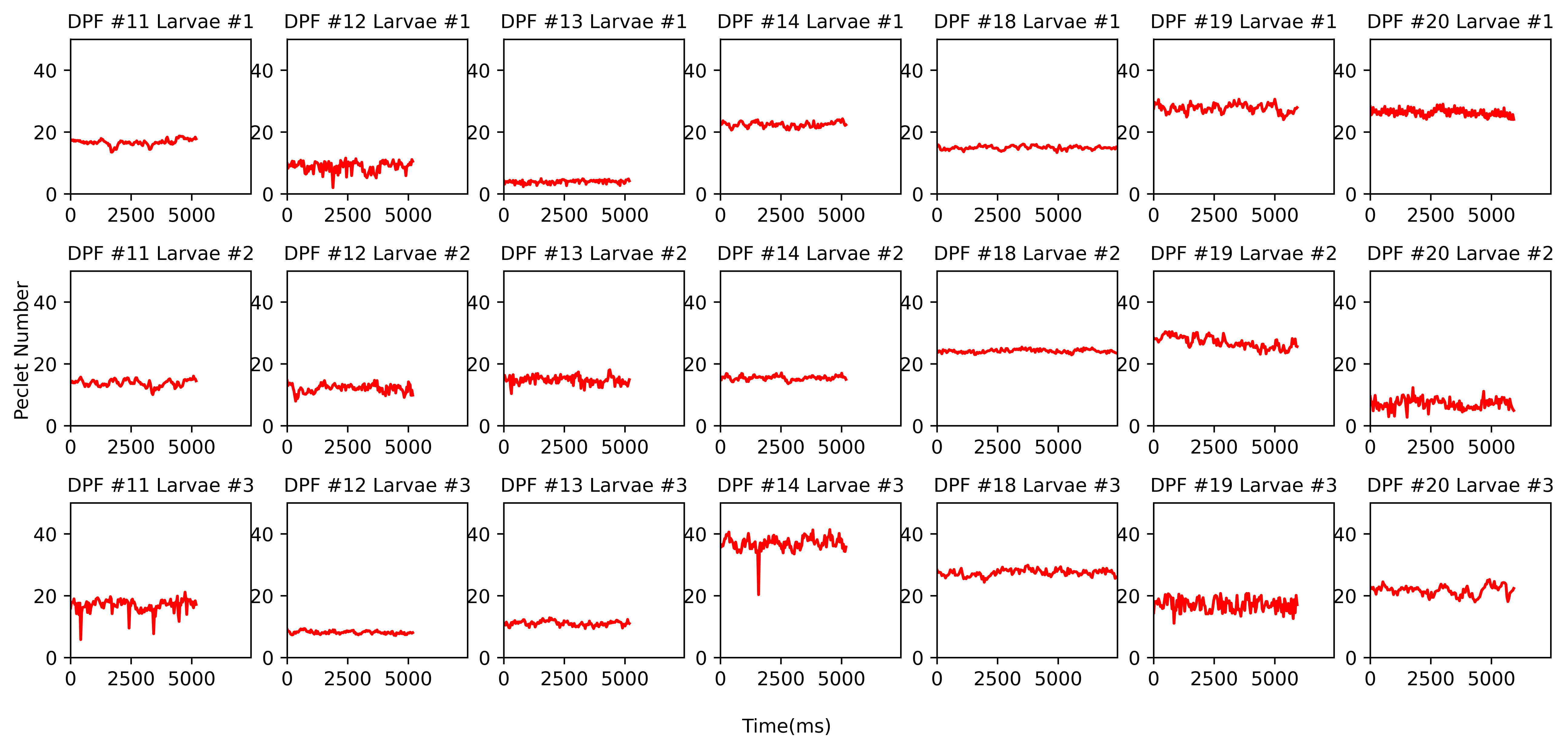}
\centering
\caption{\textbf {Supplemental Figure 5 - Péclet numbers based on average speed.} Using the speeds obtained from the ROI analysis and the larval size (estimated as a circle), Péclet number for the transport of oxygen for each larvae were calculated. Péclet numbers ranged between 10 to 40, which are similar to the spatially-informed line-of-interest analysis.}
\end{figure}

\begin{figure}[b]
\includegraphics[scale=0.55]{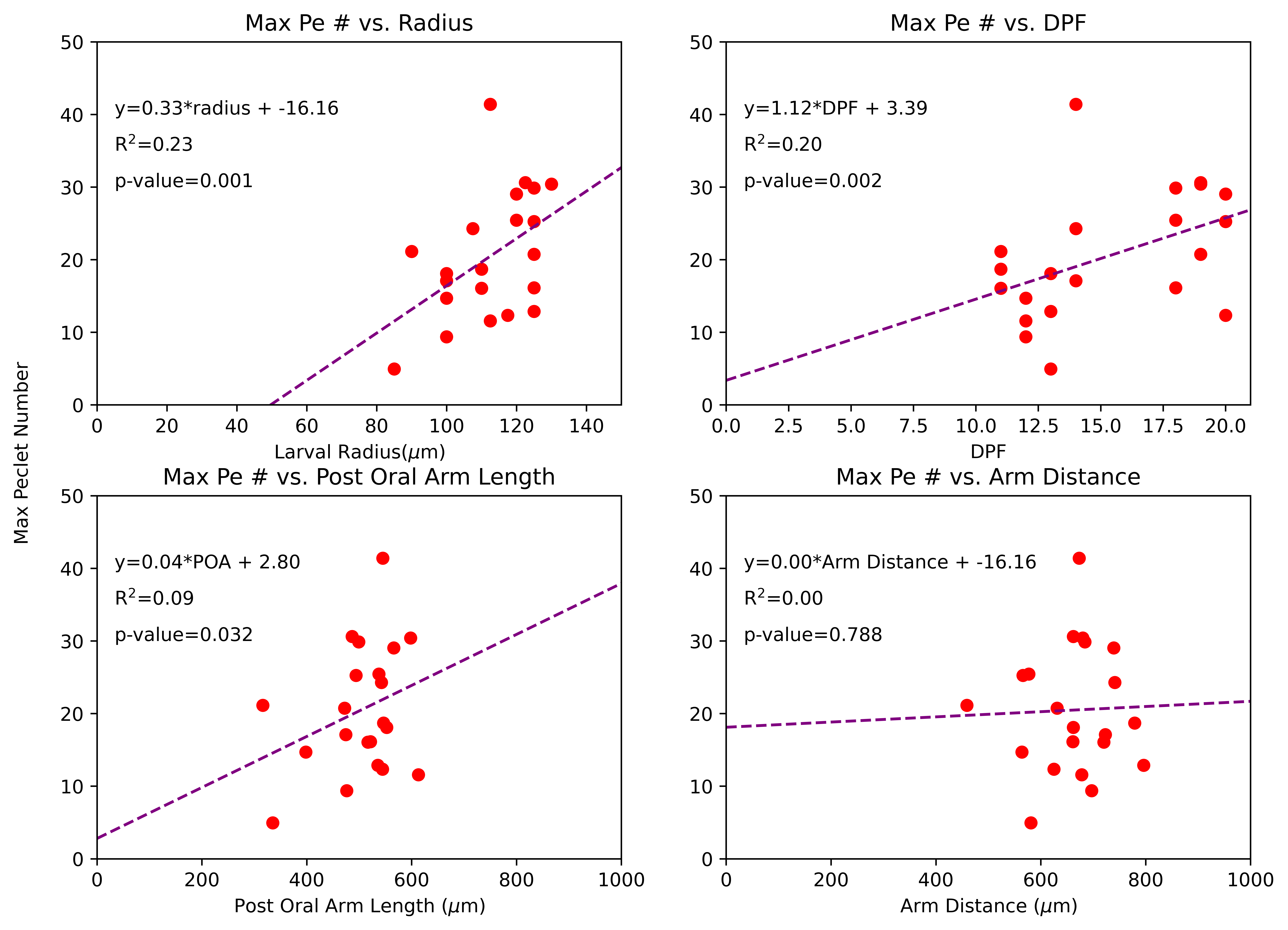}
\centering
\caption{\textbf {Supplemental Figure 6 - Relationship between Péclet numbers and larval properties.} 
Liner models were constructed to explore the relationship between the maximum Péclet number (for each larva) and larval radius, age (DPF), PO arm length, and arm distance.}
\end{figure}

\newpage

\begin{tabular}{ |p{1cm}||p{2cm}||p{3cm}||p{4cm}|| p{4cm}|   }
  \hline
  
 \multicolumn{5}{|c|}{Supplementary Table 1: Larvae and the corresponding Péclet number} \\

 \hline
    \textbf{DPF} & \textbf{Larvae} & \textbf{Diameter (microns)} & \textbf{Average Speed (mm/s)}& \textbf{Average Péclet number}\\
 \hline
 
11	&1	&220	&0.152	&18.7\\
11	&2	&220	&0.126	&16.1\\
11	&3	&180	&0.188	&21.2\\
12	&4	&225	&0.079	&11.6\\
12	&5	&200	&0.121	&14.7\\
12	&6	&200	&0.082	&9.4\\
13	&7	&170	&0.046	&4.9\\
13	&8	&200	&0.148	&18.1\\
13	&9	&250	&0.089	&12.9\\
14	&10	&215	&0.209	&24.3\\
14	&11	&200	&0.155	&17.1\\
14	&12	&225	&0.328	&41.4\\
18	&13	&250	&0.120	&16.1\\
18	&14	&240	&0.202	&25.4\\
18	&15	&250	&0.220	&29.9\\
19	&16	&245	&0.227	&30.6\\
19	&17	&260	&0.208	&30.4\\
19	&18	&250	&0.137	&20.8\\
20	&19	&240	&0.219	&29.0\\
20	&20	&235	&0.061	&12.3\\
20  &21	&250	&0.174	&25.3\\
\hline
\end{tabular}

\end{document}